\documentstyle[twoside,fleqn,espcrc2]{article}
\input{psfig}

\newcommand{\AmS}{{\protect\the\textfont2
  A\kern-.1667em\lower.5ex\hbox{M}\kern-.125emS}}

\title{
\vspace*{-0.5in}
\hspace*{5.3in}
{\normalsize UM-P-98/47}\\
\hspace*{5.3in}
{\normalsize RCHEP-98/13}\\
\vspace*{0.3in}
Evolutionary Algorithms Applied to Landau-Gauge Fixing
}

\author	{
	J.F. Markham
	\address
		{
		School of Physics, 
		University of Melbourne, 
		Vic 3052, Australia
		}
       	T. D. Kieu
	\address 
		{
		CSIRO MST, 
		Private Bag 33, 
		Clayton South MDC,
		Vic 3169, Australia
		}
	}
       
\begin{document}

\begin{abstract}
Current algorithms used to put a lattice gauge configuration into Landau
gauge either suffer from the problem of critical slowing-down or 
involve an additional computational expense to overcome it. 
Evolutionary Algorithms (EAs), which have been widely applied to other global
optimisation problems, may be of use in gauge fixing.  Also, being global,
they should not suffer from critical slowing-down as do local gradient
based algorithms. We apply EA's and also a
Steepest Descent (SD) based method to the problem of Landau Gauge Fixing and
compare their performance. 
\end{abstract}

\maketitle

\section{Motivation}
As SD is a local algorithm the time taken to converge
goes as some power ($> 1$) of $V$ the lattice volume (critical slowing down)
\cite{cucchieri96}. A solution to the problem is Fourier acceleration \cite{davies88}
but the cost of this goes as the cost of a FFT - $VlogV$ at best.
Once the gauge fixing condition has been met, Gribov copies may still exist.
These can contribute to numerical errors in the measurement of various
gauge dependent quantities,
\cite{nakamura91}
\cite{bornyakov95} 
\cite{bornyakov96} 
\cite{cucchieri97}
most notably propagators.

EA's on the other hand are global algorithms and should not suffer from critical slowing down.
They are capable of finding a global maximum and so by suitable choice of 
constraint they can in principle eliminate Gribov copies.


\section{Landau-gauge fixing on a lattice}
For Landau-gauge we seek
\begin{eqnarray}
\partial_{\mu}A^{\mu}=0
\label{eq:theta:continuum}
\end{eqnarray}
\\
A sufficient (but not necessary) condition for this to be true is the 
maximisation of
\begin{eqnarray}
F=\int d^{4}x A_{\mu}^{g}(x) A^{g\mu}(x)
\label{eq:goal:continuum}
\end{eqnarray}
where
\begin{eqnarray}
A_{\mu}^{g}(x)=A_{\mu}(x) - \partial^{\mu}\chi(x)
\end{eqnarray}
Following \cite{davies88} we do the following:
\begin{itemize}
\item Name the lattice version of $\partial_{\mu}A^{\mu}$,
 $\Theta$, and the lattice version of $F$, $F_L$. 
\item Implement SD.
\item Compare the performance of SD with a competing algorithm
in maximising $F_L$, using $\Theta$ as a measure of convergence.
\end{itemize}

\section{EA's applied to Landau-gauge fixing on a lattice}
EAs solve global optimisation problems with natural selection and genetic
style operations.
The idea is to evolve successive generations of genomes with the goal of
maximising their fitness
\cite{goldberg89} 
\cite{eareview93}
\cite{fogel95}.
We now describe the implementation of these operations to gauge fixing.

\subsection{ Coding}
\begin{figure}
\psfig{figure=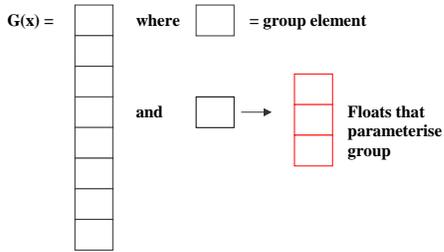,height=2.5in}
\vspace{-1.3in}
\caption{\label{figure:group_genome_pic} Encoding of groups }
\end{figure}

We encode the gauge transformations $G(x)$ ({\em not} the gauge fields $U(x)$). 
Each genome is made of group elements for each site of the lattice 
(1D is shown in Figure~\ref{figure:group_genome_pic}).
Each group element is made of floats which represent the angles parameterising it.
\begin{itemize}
\item U(1) - just one angle.
\item SU(2) - use the angles in O(3).
\item SU(3) factorise into 3 SU(2) matrices acting on the three SU(2) subspaces in SU(3).
\end{itemize}

\subsection{Mutation}
The floats are mutated by adding some normally distributed random numbers to them.
The standard deviation of the distribution can vary according to the generation, 
genome and position within the genome.

\subsection{Recombination}
Cutting and splicing of the genome as used in Ref.~\cite{goldberg89} can be used. 
Best results are achieved by cutting on group boundaries and by 
having the genome duplicate the dimensionality and boundary conditions of 
the lattice. More successful methods are those which globally interpolate between 
genomes. 

\subsection{Fitness}
\begin{figure}
\psfig{figure=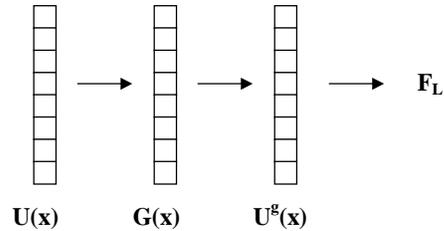,height=2.5in}
\vspace{-1.5in}
\caption{\label{figure:fit_genome_pic} Evaluating the fitness of a gauge transformation }
\end{figure}

The fitness of a genome is found by applying the $G(x)$ that it represents 
to the original gauge fields and then evaluating $F_L$ on the result  
as shown in Figure~\ref{figure:fit_genome_pic}.

Genomes are selected from a population for breeding using algorithms
which favour those with high fitness while at the same time maintaining
diversity in the population. The last condition is to allow the population
to escape local maxima.

\section{Typical EA results}
\begin{itemize}
\item The graphs that follow are not measures of flops.
\item One generation of EA takes more time than one iteration of SD.
\item There is also the issue of memory. For an EA one must store many genomes, each of which
has the size of a gauge transformation for the entire lattice.
\item All lattices used are $8^1$ with $U(1)$ gauge fields with the exception of 
the last two figures which are $16^1$ with $U(1)$ gauge fields. 
\end{itemize}

\begin{figure}
\psfig{figure=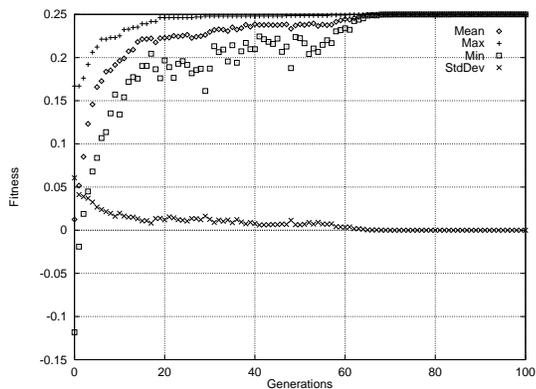,height=2.0in}
\vspace{-.2in}
\caption{\label{figure:nocrossga} Fitness of Population vs Generations a typical plot for EA }
\end{figure}

\begin{figure}
\psfig{figure=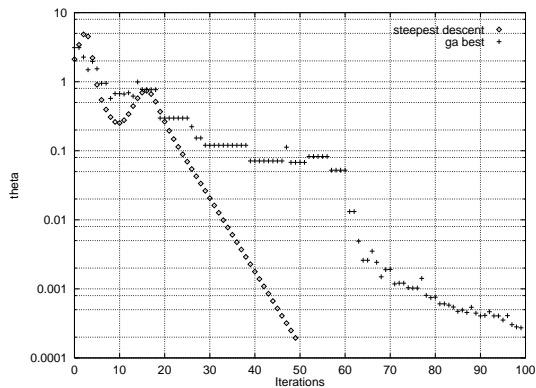,height=2.0in}
\vspace{-.2in}
\caption{\label{figure:nocrosstheta} $\Theta$ vs Generations - a comparison of the two methods for a typical run }
\end{figure}

Figure~\ref{figure:nocrossga} shows the statistics of a population as the algorithm progresses.
The standard deviation of the fitness gives a measure of convergence (in addition to $\Theta$).
Figure~\ref{figure:nocrosstheta} shows a direct comparison between SD and EA. On average
SD performs better, and we have always found this to be the case.

\begin{figure}
\psfig{figure=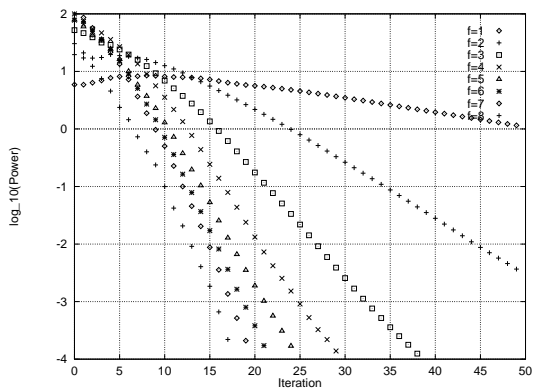,height=2.0in}
\vspace{-.2in}
\caption{\label{figure:sdspec} Power vs Iterations - Fourier components of successive gauged fields using SD }
\end{figure}

\begin{figure}
\psfig{figure=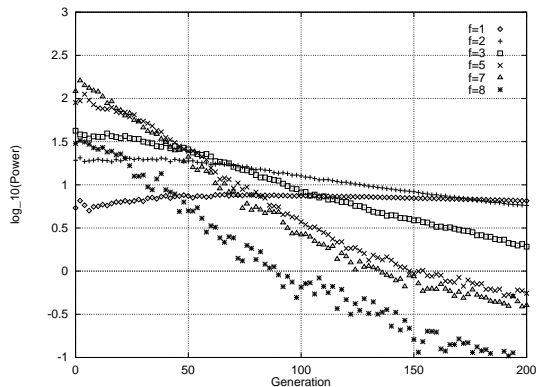,height=2.0in}
\vspace{-.2in}
\caption{\label{figure:gaspec} Power vs Generations - Fourier components of successive gauged fields using EA }
\end{figure}
The reason for this might be that there is no
gradient information available to EA's. Close to convergence the genomes
are doing a random walk (ie. without guidance) in the phase space of gauge transformations. 
Surprisingly, spectral decomposition of the gauged configurations suggests  
that the EA's are suffering from a similar problem that afflicts SD \cite{davies88}.
Examination of Figures~\ref{figure:sdspec} and \ref{figure:gaspec} shows that the slow decay
of low momenta modes in both algorithms is a problem.

\section{Conclusion}
In this preliminary study, our efforts have concentrated on improving the 
rate of convergence of EAs. This seems to go as some power 
of $r/n$ where $r$ is the distance from the actual maximum and 
$n$ is the number of floats in a genome.
Presently we are looking at their application to distinguishing between Gribov copies as 
in theory EAs should be able to find the global maximum effectively.  Various hybrid methods 
combining EAs and local gradient methods -- such as SD or its variations -- are also being 
considered.

\end{document}